\documentclass[twocolumn,showpacs,preprintnumbers,floatfix,prl]{revtex4}
\usepackage[usenames]{color}
\usepackage{graphicx}
\usepackage{amsmath}

\begin{document}

\title{Compaction of anisotropic granular materials : \\
experiments and simulations}

\author{G. Lumay and N. Vandewalle}

\affiliation{GRASP, Institut de Physique B5, Universit\'e de Li\`ege, \\ B-4000 Li\`ege, Belgium.}

\begin{abstract}

We present both experimental and numerical investigations of compaction in granular materials composed of rods.  As a function of the aspect ratio of the particles, we have observed large variations of the asymptotic packing volume fraction in vertical tubes. The relevant parameter is the ratio between the rod length $\ell$ and the tube diameter $D$. Even the compaction dynamics remains unchanged for various particle lengths, a 3d/2d phase transition for grain orientations is observed for $\ell/D = 1$. A toy model for the compaction of needles on a lattice is also proposed. This toy model gives a complementary view of our experimental results and leads to behaviors similar to experimental ones.

\pacs{05.70.Jk,45.70.Cc}
\end{abstract}

\maketitle

\section{Introduction}

Granular matter has been the subject of numerous studies since the last decade \cite{deGenne,granularSolidLiquidsGases,Duran,Kudrolli}. Indeed, most of the industrial products are processed, transported and stocked in a granular state. The packing density of those granular materials becomes therefore a relevant parameter for a broad range of applications. The best way to reduce the costs for the manipulation of such granular materials is to increase the packing density $\rho$. This can be achieved by tapping or vibrating the vessel containing the grains. 

Various experimental studies \cite{densFluc,densityRelax} have underlined the fact that the dynamics of compaction is a complex problem. The compaction dynamics is indeed characterized by a slow dynamics \cite{densityRelax}. Different laws have been proposed for the volume fraction evolution of a granular material as a function of the number $n$ of taps. Among others, one has proposed the inverse logarithmic law 
\begin{equation}
\label{eq:empirical}
\rho(n) = \rho_{\infty} - \frac{\Delta \rho}{1 + B \; \ln(1+\frac{n}{\tau})}
\end{equation} where the parameters $\rho_{\infty}$ and $\Delta \rho$ are respectively the asymptotic volume fraction and the maximum variation of the volume fraction. The dimensionless parameter B depends on the acceleration during each tap and $\tau$ is the relaxation time of the reorganization process. This inverse logarithmic law of $n$ was obtained in numerical models for compaction like the Tetris model and could also be derived from theoretical arguments \cite{slowRelax}.

The great majority of earlier experiments have considered the compaction dynamics of spherical grains \cite{pierrePhilippe,simusphere}. Only a few papers discuss the problem of anisotropic particles \cite{rods,prolatePacking}. It has been reported that the compaction dynamics exhibits different regimes associated to respectively grain translations and grain rotations. 

It has been also reported \cite{thin-rod,spherocylinders} that the random packing density of spherocylinders or rods decreases when the aspect ratio $\ell/d$ of these objects increases.

In this paper, we present an experimental study of compaction dynamics for cylindrical particles in a vertical tubes (Figure \ref{im:expSetup}). We investigate the volume fraction as a function of the aspect ratio of the grains $\ell/d$ and as function of the ratio between the grain length and the tube diameter $\ell/D$. We will see that according to the ratio $\ell/D$, restrictions on grain orientations appear. These restrictions have large influence on the asymptotic volume fraction.

A toy model for compaction is also proposed for the case of anisotropic grains.

\section{Experimental study}

\subsection{Experimental set-up}

\begin{figure}[h]
\begin{center}
\includegraphics[scale=0.25]{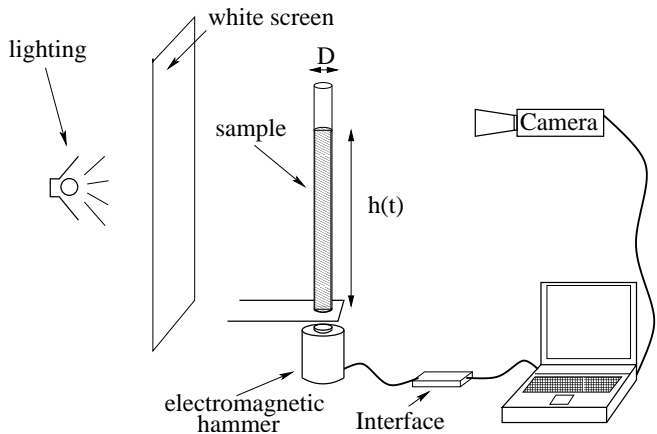} 
\end{center}
\caption{Sketch of our experimental setup. The granular material is placed in a vertical glass tube. Taps are generated by an electromagnetic hammer controlled by a computer. A camera records the granular/air interface after each tap.}
\label{im:expSetup}
\end{figure}

Our experimental setup for the study of compaction is different from usual ones. Indeed, an oscillating system is generally used to produce the compaction process. In our experiment, an electromagnetic hammer, which is controlled by a computer via an interface, is placed below a vertical tube for tapping. Thus, we act on the system with a series of intense and short taps. The intensity, the number and the frequency of the taps can be controlled. A CCD camera records the tube during the whole experiment. Therefore, we can measure the evolution of the height $h$ of the granular/air interface by image analysis. A ruler is placed along the tube in order to calibrate the measurements. The picture analysis for the volume fraction measurements is made by an algorithm especially developed for this task. On the pictures, grains appear in dark and the rest of the system appears in bright. An average of the pixel values is performed on the vertical of the picture. We obtain an abrupt variation of the average at the position of the granular/air interface. The maximum of the first derivative gives the position of the interface. The volume fraction $\rho$ is estimated through the value of $h$, the density of grains and the weight of the entire column. One should note that in our experiment, we are mainly interested on the volume fraction averaged over the entire granular column. A sketch of our setup is given in Figure \ref{im:expSetup}.

The granular material we used is made of identical cylinders. The diameter $d$ of these cylinders is 1.2 mm and their length varies between 2 mm and 26 mm. Their aspect ratio $\ell/d$ varies between 1.6 and 21. The cylinders are placed in a vertical glass tube. The internal diameter $D$ of the glass tube can vary between 8 mm and 24 mm.

For the preparation of the loose packing, grains are poured in the glass tube with a funnel before each run. We have also tested a decompaction by inflating air from the bottom of the tube. This does not change our results obtained with the rain method.

In order to obtain a saturation of the volume fraction when tapping, a minimum of 2000 taps was applied before stopping each experiment.

Before investigating the compaction dynamics, let us present some additional information about tapping. We have measured the acceleration experienced by the bottom of the glass tube after each tap. The analysis of a single tap with an ultrafast camera gives a typical acceleration of 12 $g$ during a short period of 2 ms. The amplitude of the vertical motion is approximately 0.25 mm. We have also measured the relaxation time of the system after a tap (the time after which there is no movement in the tube). We obtained at most 0.12 s. In all our experiments, the successive taps are always separated by at least 0.2 seconds. This time interval is long enough to avoid any overlap between relaxation mechanisms provoked by two successive taps.

\subsection{Compaction dynamics}

\begin{figure}[htbp]
\begin{center}
\includegraphics[scale=0.7]{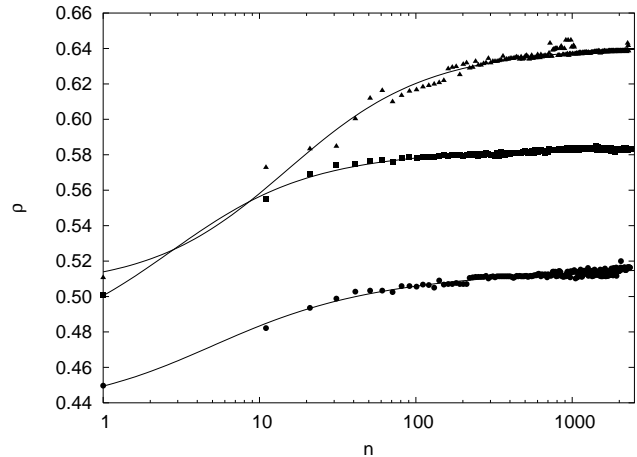} 
\end{center}
\caption{Experimental compaction curves : volume fraction $\rho$ as a function of the tap number $n$. The curves are fits using Eq. (\ref{eq:empirical}). Three different cases are illustrated : $\ell$ = 4 mm (squares), $\ell$ = 10 mm (circles) and $\ell$ = 20 mm (triangles) for D = 10 mm.}
\label{im:expCompCurve}
\end{figure}

We have analysed the compaction for different tube diameters $D$ and for different lengths $\ell$ of the rods. The figure \ref{im:expCompCurve} presents typical curves of compaction for three different lengths. Each curve is well fitted by the inverse logarithmic law of  Eq(\ref{eq:empirical}). Curves obtained in other glass tubes present similar behaviors. We do not observe different stages in the compaction process as in \cite{rods}. In fact, several stages have been observed for a vertical acceleration less than $\Gamma = 7.5$. In our experiment, the dimensionless acceleration is much larger ($\Gamma = 12$) as underlined here-before. 

\begin{figure}[htbp]
\begin{center}
\includegraphics[scale=0.7]{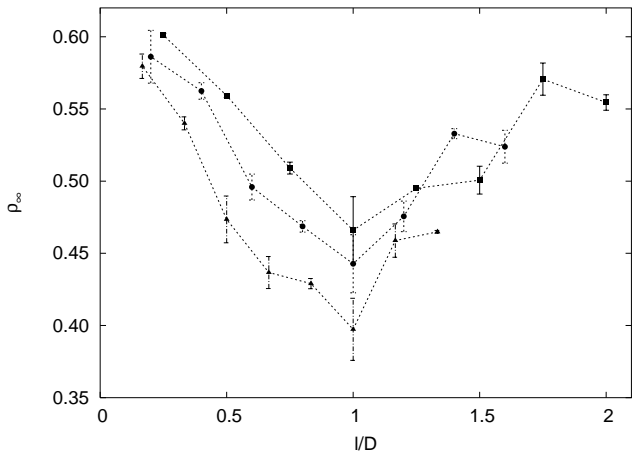}
\end{center}
\caption{Asymptotic volume fraction $\rho_{\infty}$ as a function of the ratio $\ell/D$ for three different tube diameters : $D$= 8 mm (squares), $D$ = 10 mm (circles) and  $D$ = 12 mm (triangles). Error bars are indicated.}
\label{im:expDensFin}
\end{figure}

All experimental compaction curves have been fitted with the inverse logarithmic law, i.e. with Eq.(1). From the fits, a relevant parameter is the asymptotic volume fraction $\rho_{\infty}$. The figure \ref{im:expDensFin} shows the asymptotic volume fraction $\rho_{\infty}$  as a function of the ratio $\ell/D$ for different tube diameters. The asymptotic volume fraction presents large variations when the ratio $\ell/D$ is changed. In all experiments, we observe a minimum of $\rho_\infty$ when the cylinder length is equal to the tube diameter, i.e. when $\ell = D$. From these results, one could ask whether the compaction dynamics changes or not when varying $\ell/D$.

\begin{figure}[htbp]
\begin{center}
\includegraphics[scale=0.7]{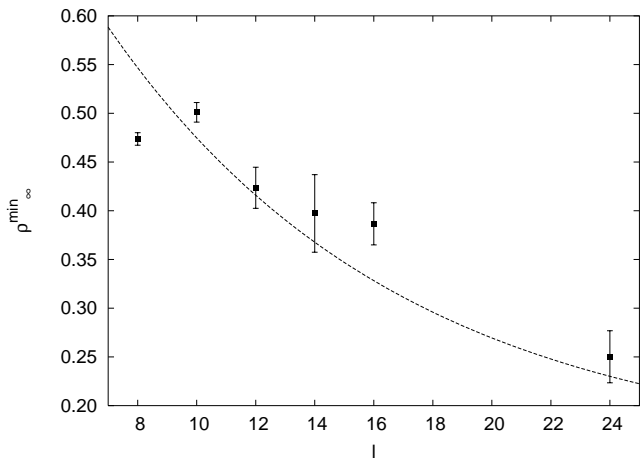}
\end{center}
\caption{Asymptotic volume fraction $\rho_{\infty}^{min}$ as a function of $\ell$ for the particular case $\ell =D$. Error bars are indicated. The curve is only a guide for the eyes}
\label{im:ExpDensFinlEqD}
\end{figure}

While the asymptotic volume fraction of the packing does not vary much with $D$ for both extreme cases $\ell/D >> 1$ and $\ell/D << 1$, the situation is quite different for the particular case $\ell = D$. In that case, the asymptotic volume fraction $\rho_{\infty}$ decreases when the tube diameter $D$ increases. Figure \ref{im:ExpDensFinlEqD} is a plot of $\rho_{\infty}^{min}$ as a function of $l$ for the particular case $\ell = D$. The line in Figure \ref{im:ExpDensFinlEqD} is only a guide for the eyes. 

\subsection{Discussion}

As for common liquid cristals, the rods have two parameters for 
describing a possible long-range order : position and orientation. 
Each of them plays a role in the packing fraction. Since random 
positions are usually observed even for spherical grains, the largest 
source of heterogeneities in our experiment comes from grain 
orientations. The orientation disorder is reduced when $\ell > D$. 
One should also note that the vertical configurations which are 
favored by the geometrical constraints do not correspond to energy 
minimization. Indeed, the potential energy for a grain is minimized 
when this grain is horizontal.

\begin{figure}[htbp]
\begin{center}
\includegraphics[scale=0.5]{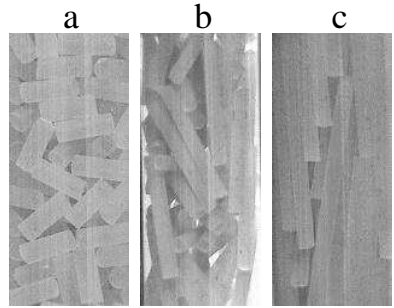}
\end{center}
\caption{Pictures of the tube after 2000 taps and for a tube diameter $D=$ 10 mm. Three different cases are illustrated: (a) $\ell$ = 4 mm, (b) $\ell$ = 10  mm, (c) $\ell$ = 20 mm.}
\label{im:CompDiffGrainsExp}
\end{figure}

Figure \ref{im:CompDiffGrainsExp} presents typical pictures of the 
packing for various ratios $\ell/D$. The pictures were taken after 
several taps. From the observation of the left picture shown in 
Figure \ref{im:CompDiffGrainsExp}, one understands that all grain 
orientations are possible for $\ell<D$. In that case, the ideal 
packing could be realized in various directions. For $\ell>D$, 
horizontal orientations are forbidden. The grains are ideally 
arranged along the tube direction, i.e. along the vertical direction. 
This global ordering of the grains is well seen on the right picture 
of Figure \ref{im:CompDiffGrainsExp}. When $\ell=D$, a few grains are 
placed in the horizontal direction. Large voids can be seen near 
horizontal grains. This case is shown in the central picture of 
Figure \ref{im:CompDiffGrainsExp}.

The large variation of the packing fraction, which is observed when 
the needle length $\ell$ is equivalent to the tube diameter $D$, 
should be attributed to a 3d-2d transition. Let us discuss some 
particular cases. For $\ell \ll D$, the situation tends to a packing 
of isotropic objects. The 3d ideal volume fraction of identical 
spheres is $\rho= \pi / 3 \sqrt{2} \approx 0.74$ while the random 
close packed limit is $\rho \approx 0.58$, i.e. a value close to what 
we have measured for the smaller grains. In the case of long needles 
($\ell \gg D$), all needles have a vertical orientation. The ideal 
volume fraction is given by the fraction of disks arranged along a 
hexagonal packing in the plane crossing the glass tube, i.e. $\rho = 
\pi \sqrt{3} /6 \approx 0.91$. Of course, this ideal ordered packing 
is never realized. The random case corresponds to a packing fraction 
$\rho \approx 0.82$ which is the upper limit of our experimental 
measurements.

From the above discussion on both extreme cases, one expects a global increase of the volume fraction from 0.58 (random sphere packing) to 0.82 (random disk packing) when the ratio $\ell/D$ increases. How to explain that volume fraction presents a minimum for $\ell \approx D$? This minimum is the consequence of the competition between two phenomenons. First, the packing density decreases when the aspect ratio of grain $\ell/d$ increases \cite{thin-rod,spherocylinders}. Secondly, we observe an arrangement of grains along the tube direction when $\ell/D > 1$. This arrangement leads to an increase of the packing density. The first phenomenon explains the decrease of $\rho_{\infty}$ for $\ell/D$ situated between 0.2 and 1. The second phenomenon explains the increase of $\rho_{\infty}$ for $\ell/D$ situated between 1 and 2. Furthermore, when $\ell \approx D$, some horizontal grains can block the vertical motion of the other grains and form jams.

In Figure \ref{im:ExpDensFinlEqD} the minimum value of $\rho_{\infty}$ is found to decrease with $\ell$ (=$D$). This decrease is explained by the increase of the aspect ratio $\ell/d$ and by the presence of jams when a grain is horizontal.

\section{SIMULATIONS}

\subsection{Toy model}

In order to reproduce our experimental results, we propose a toy model on a 2d lattice. Grains are needles having different orientations. They are placed on a square lattice. Each grain has 8 different possible orientations which are illustrated in Figure 7. Lattice boundaries are closed. The length $\ell$ of the needles is a parameter to be investigated. The width $W$ of the lattice is the second parameter of the toy model. 

\begin{figure}[htbp]
\begin{center}
\includegraphics[scale=0.3]{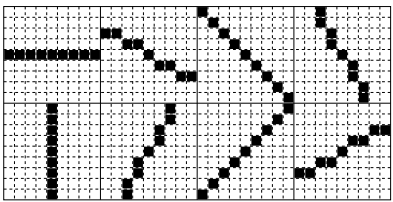} 
\end{center}
\caption{Eight possible orientations of a needle with $\ell = 9$ on a square lattice.}
\label{im:Grain}
\end{figure}

The first step of the model consists in filling the lattice with anisotropic grains. The first configuration is obtained by the rain method. Grains are dropped sequentially on the top of the lattice. Both grain position and grain orientation are chosen randomly. The grains follow a vertical motion until they touch another particle or the lattice edge. 

Once the lattice is filled, the model consists in applying a series of taps. The procedure to simulate a single tap is the following. We randomly choose some grains as many times as the number of grains in the system. Thus, every grain in the system has a chance to move. For a selected grain, we have the choice between two motions :  rotation or translation. The selected motion for that grain has also two possible directions. Each motion and each direction is chosen randomly with a probability $0.5$. The operation is executed only if it does not lead to some overlap of the grains. The volume fraction $\rho$ of the packing is measured after every tap. It should be also underlined that this toy model does not consider  the gravity when randomly selecting new orientations for the needles. Indeed, the effect of gravity becomes negligible when $\rho$ becomes large (after a few taps). 

\begin{figure}[htbp]
\begin{center}
\includegraphics[scale=0.5]{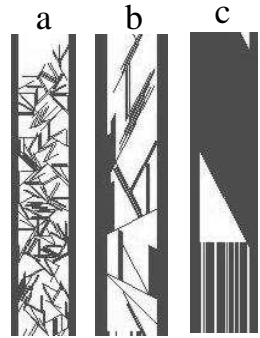}
\end{center}
\caption{Typical simulated packings in a 2d system. Three different needles ($\ell=13,41,77$) are illustrated for the same lattice ($W=41$). Different heterogeneities from disorder to long-range order can be seen. Large voids are obtained when $\ell = W$.}
\label{im:CompDiffGrainsSimu}
\end{figure}

\subsection{Numerical results}

\begin{figure}[htbp]
\begin{center}
\includegraphics[scale=0.7]{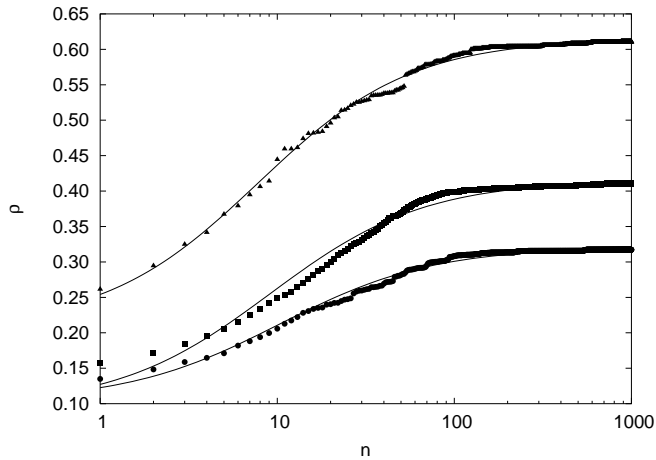}
\end{center}
\caption{Simulated compaction curves: volume fraction $\rho$ as a function of the tap number $n$. The curves are fits using the inverse logarithmic law Eq. (\ref{eq:empirical}). Three different cases are illustrated : $\ell$ = 9 (squares), $\ell$ = 21 (circles) and $\ell$ = 41 (triangles). All data are obtained for $W$ = 21.}\label{im:SimuCompCurve}
\end{figure}

\begin{figure}[htbp]
\begin{center}
\includegraphics[scale=0.7]{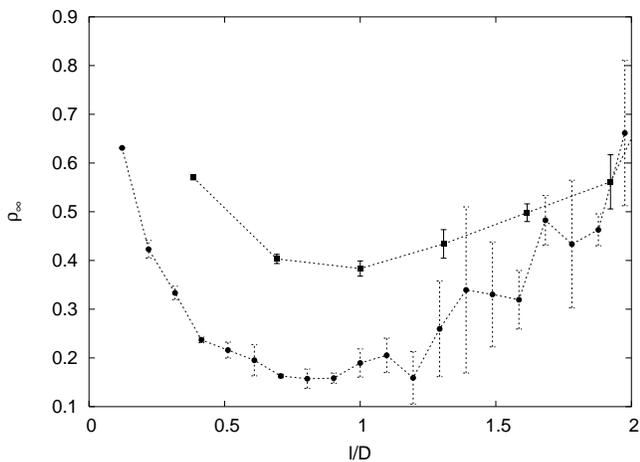}
\end{center}
\caption{Asymptotic volume fraction $\rho_{\infty}$ as a function of the ratio $\ell/D$ for different lattices: $W$=13 pixels (squares) and $W$=41 pixels (circles). Error bars are indicated.}
\label{im:SimuDensFin}
\end{figure}

In the simulations, we have observed the evolution of the volume fraction $\rho$ as a function of the tap number $n$. Figure \ref{im:SimuCompCurve} presents typical compaction curves obtained numerically for 3 different situations. An inverse logarithmic dynamics has been found in all cases. From the fits using Eq.(1), we have obtained the evolution of the saturation density $\rho_{\infty}$ for various ratios $\ell/W$ and for different lattices. The results are displayed in Figure \ref{im:SimuDensFin}. As for experiment the volume fraction presents a minimum when $\ell = W$. 

\begin{figure}[htbp]
\begin{center}
\includegraphics[scale=0.7]{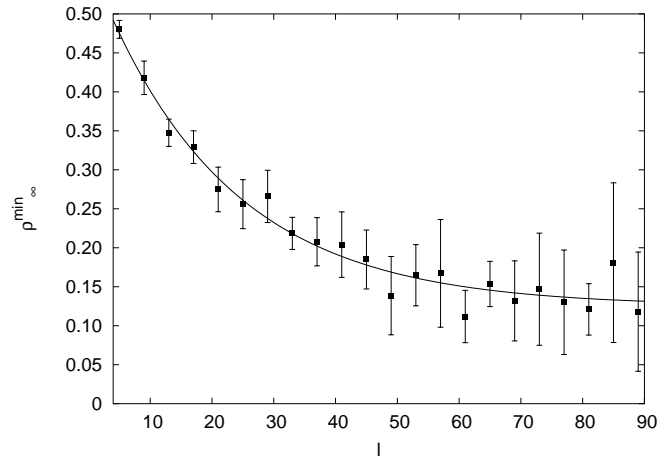}
\end{center}
\caption{Asymptotic volume fraction $\rho_{\infty}$ as a function of $\ell$ for the particular case $\ell =W$. Error bars are indicated. The curve is only a guide for the eyes}
\label{im:SimuDensFinlEqD}
\end{figure}

Moreover, we have numerically investigated the particular case $\ell=W$.  Figure \ref{im:SimuDensFinlEqD} presents the volume fraction $\rho_\infty$ as a function of $W$ in a semi-log plot. The volume fraction is seen to decrease exponentially with the lattice width $W$.  One has
\begin{equation}
\rho_\infty = \rho_R - A \; exp(-\frac{W}{L})
\end{equation} with $\rho_R = 0.125 \pm 0.009$ which is a residual density for a jammed phase and $L = 21.1 \pm 2.3$ is a characteristic length of particle.

\subsection{Discussion}

Our toy model reproduces qualitatively our experimental results. 
Indeed, the compaction curves look similar to the experimental curve 
(see figure \ref{im:SimuCompCurve}). A minimum of the volume fraction is also observed 
when the needle length $\ell$ becomes equivalent to the width $W$ of 
the lattice. The exponential scaling of the minimum (figure \ref{im:SimuDensFinlEqD}) is unexpected and needs further theoretical studies.

The large variation of the packing fraction, which is observed when 
the needle length $\ell$ is equivalent to the width $W$ of 
the lattice, should be attributed to a 2d-1d transition. Figure \ref{im:CompDiffGrainsSimu} presents typical packings on the same lattice. Three 
different lengths $\ell$ are illustrated: $\ell < W$, $\ell = W$ and 
$\ell > W$. In the former case, the grains have various orientations. 
Small voids can be seen. In the second case, a few grains have 
horizontal orientations blocking the lattice. This jammed situation 
could create large voids. In the third case, a large majority of the 
grains have a vertical orientation and packs ideally along the 
lattice. A few diagonal grains creates voids in the lattice.

\section{Conclusion}

In this paper, we present an experimental study of compaction in granular materials composed of anisotropic particles.  As a function of the aspect ratio of the particles, we have observed large variations of the asymptotic packing volume fraction in vertical tubes. We have observed a transition 3d/2d when the needle length becomes equal to the width of the container. A toy model is proposed and reproduces numerically what we observed.

\section{Acknowledgements}

This work has been supported by the contract ARC 02/07-293. The authors thank H.Caps, S.Dorbolo, F.Ludewig, M.Prochnow and S.Theis for valuable discussions.

\end{document}